\documentclass[aps,prb,reprint,showpacs,superscriptaddress,groupedaddress]{revtex4-1}
\usepackage{graphicx}
\usepackage{amssymb}
\usepackage{dcolumn}
\usepackage{latexsym}
\usepackage{rotating}
\usepackage[usenames,dvipsnames]{xcolor}
\usepackage{float}
\usepackage{epsfig}
\usepackage{psfrag}
\usepackage{bm}
\usepackage{eucal}
\usepackage{braket}
\usepackage{enumerate}
\usepackage{longtable}
\usepackage{subfigure}
\usepackage{bm}
\usepackage{hyperref}
\usepackage{amsfonts}
\usepackage{dcolumn}
\usepackage{bm}
\usepackage{subfigure}
\newcommand{\be}{\begin{equation}}
\newcommand{\ee}{\end{equation}}
\newcommand{\bn}{\begin{eqnarray}}
\newcommand{\en}{\end{eqnarray}}

\usepackage{color} 

\usepackage{hyperref}
\begin{document}

\author{S. Koley$^{1}$}\email{sudiptakoley20@gmail.com}
\author{Saurabh Basu$^{2}$}
\title{Orbital Selectivity and Magnetic Ordering in Fe intercalated Dirac Semimetal Bi$_2$Se$_3$}
\affiliation{$^{1}$ Department of Physics, North Eastern Hill University, 
Shillong, Meghalaya, 793022 India}
\affiliation{$^{2}$ Department of Physics, Indian Institute of Technology Guwahati, Assam, 781039 India}
\begin{abstract}
\noindent In this paper we investigate the intercalation effects of Iron (Fe) in the
van der Waals gap of Bi$_2$Se$_3$ on the magnetic and transport properties using first-principles 
band structure estimations combined with dynamical mean-field theory. 
The Dirac cone in the band structure of parent Bismuth Selenide is modified 
via Fe intercalation at moderate densities.
Further inclusion of
electronic correlations found to result in the emergence of novel and exotic
properties in an intercalated Bi$_2$Se$_3$. Accompanied by unconventional 
structural effects, the onset of an 
orbital selective metal insulator transition in the Fe 3$d$ orbitals 
brings about a magnetic phase transition in the Fe intercalated Bi$_2$Se$_3$. 
Additionally we have explored the dependency of the 
electron-electron correlations on the magnetic ordering
and the effects of intercalation in establishing new physical properties.
\end{abstract}

\maketitle
\section{INTRODUCTION}

Topological insulators (TI) denote materials with properties dominated
by the bulk insulating states, while their surface electronic
structures are found to be metallic in nature\cite{kane,mele,bernevig,konig}. In these TIs, if  
spin-orbit coupling (SOC) exceeds the band gap and induce a band inversion, 
a Dirac type band structure emerges, and subsequently, the conducting states appear at 
the surfaces as demanded by the time-reversal symmetry.
In recent times, these topological surface states are engineered in numerous
ways to inspect their exotic properties and applicability\cite{srxbi}. 
Specifically, intercalation and doping by the transition metals (TM) in 
topological insulators and breaking of the time reversal symmetry are new features 
that are gathering interest in recent times. In addition, several exciting phenomena such as,
the realization of quantum anomalous Hall effect\cite{yu,chang} etc. are likely to be possible.

Intercalation by foreign materials into the layered topological insulators changes
properties of the parent compounds and possesses versatile applications, such as in superconductors, 
ambipolar transistors, quantum computers, battery electrodes and solid 
lubricants\cite{wang}. In particular, intercalation by the transition 
metals are expected to modify the time reversal symmetry induced band inversion and change the linearly
dispersing bands by the reformed topological surface states. A well-known three
dimensional topological insulator is Bi$_2$Se$_3$\cite{qi} with a 0.3 $eV$ gap in its bulk,
alongwith the presence of a Dirac cone in the $K-\Gamma-M$ direction.
Ab initio electronic structure calculations \cite{zhang} and electron scattering 
experiments have convincingly demonstrated topological properties of Bi$_2$Se$_3$.
Further, doped Bi$_2$Se$_3$ is currently being investigated due to the prospects of
observing a new phenomenon that occurs when the topological surface states interact with different 
type of impurities or with other electronic states in the bulk. Intercalating Bismuth
Selenide with Cu to get superconductivity below ∼ 4K was a potential discovery\cite{hor}.
Since it is a layered material, there is a van der Waals gap (gap between two 
consecutive layers of the material and the layers have van der Waals interaction 
between them) in the 
crystal structure, making it favourable for intercalation with other materials as well, such as, 
Fe, Sr, Ag and many others. Intercalation into host materials with these  
has the prospects of achieving new energy storage materials.

The present paper focuses on the intercalation of topological insulating compound Bi$_2$Se$_3$ with Fe. 
The percentage of intercalation in these layered materials is 
determined by the physical size of the inserted material, structural stability and 
energetically favourable state of the host after insertion. Most of the time, 
the intercalation is governed by the likelihood of the host to retain
its charge neutrality\cite{koski}. Intercalation with small alkali metals, such as, Li 
etc. is relatively easy due to the size, while the ionic nature of the intercalant determines 
the probability of success of the intercalation in most cases. With all these issues in mind, 
intercalation with a zerovalent material is of a greater advantage, 
since it does not change the oxidation state of the host material, which, as
a result, allows the insertion of a lot of zerovalent materials 
(e.g., Au, Ag, Fe, Cu and Ni) in the topological insulators\cite{koski}. It does not 
affect the layered host Bi$_2$Se$_3$, thus enabling an accurate 
intercalation. Here we used nearly 10 atomic percent of zerovalent Fe into layered Bi$_2$Se$_3$ 
crystals. Earlier studies have observed that Fe-doped 
Bi$_2$Se$_3$ is dominated by ferromagnetic interactions\cite{salman}. It is relevant to mention here
that magnetism can also be achieved by Cr doping, where onset of antiferromagnetic correlations was observed \cite{jung}.
Further quantum anomalous Hall state was also discovered in vanadium doped topologically
insulating films\cite{changnmat}, which makes magnetic topological insulators as candidates for
electronic applications. The discovery of the magnetic topological 
insulators has become interesting with an appearance of integer moments which 
may be related to half-metallic behaviour \cite{larson}.

Parent Bi$_2$Se$_3$, when doped with TM, such as, Cr and Fe, shows an
insulating behaviour with reduced band gaps due to the hybridization 
between the $d$ orbitals of the TM with the Se-$p$ orbitals \cite{zhang}.
Interestingly, in the TI systems, strong spin-orbit coupling results in 
different parity valance, and the conduction bands cross, thereby   
opening up a band gap to produce an inverted band structure. 
However in an intercalated Bi$_2$Se$_3$, the TM atoms cause disorder which often 
induce some impurity bands in the band gap, leading to a number of versatile properties, 
including, in some cases, the strength of the spin-orbit coupling is reduced. It is 
well known that the 3$d$ electrons in TM atoms contribute to magnetism with a 
maximum of five unpaired electrons. These 3$d$ electrons may easily hybridize with the
surrounding TI atoms, thereby reducing their average moments. For these reasons, several
experimental explorations have emerged on magnetically doped TIs\cite{zheng}. 
However on the theoretical front, research on Fe 
intercalation in Bi$_2$Se$_3$ is still lacking. Therefore, there is a strong 
motivation for us to study the effects of Fe intercalation on structural and 
electronic properties of Bi$_2$Se$_3$.    

\begin{figure}
\centering
(a)
\includegraphics[angle=0,width=0.41\columnwidth]{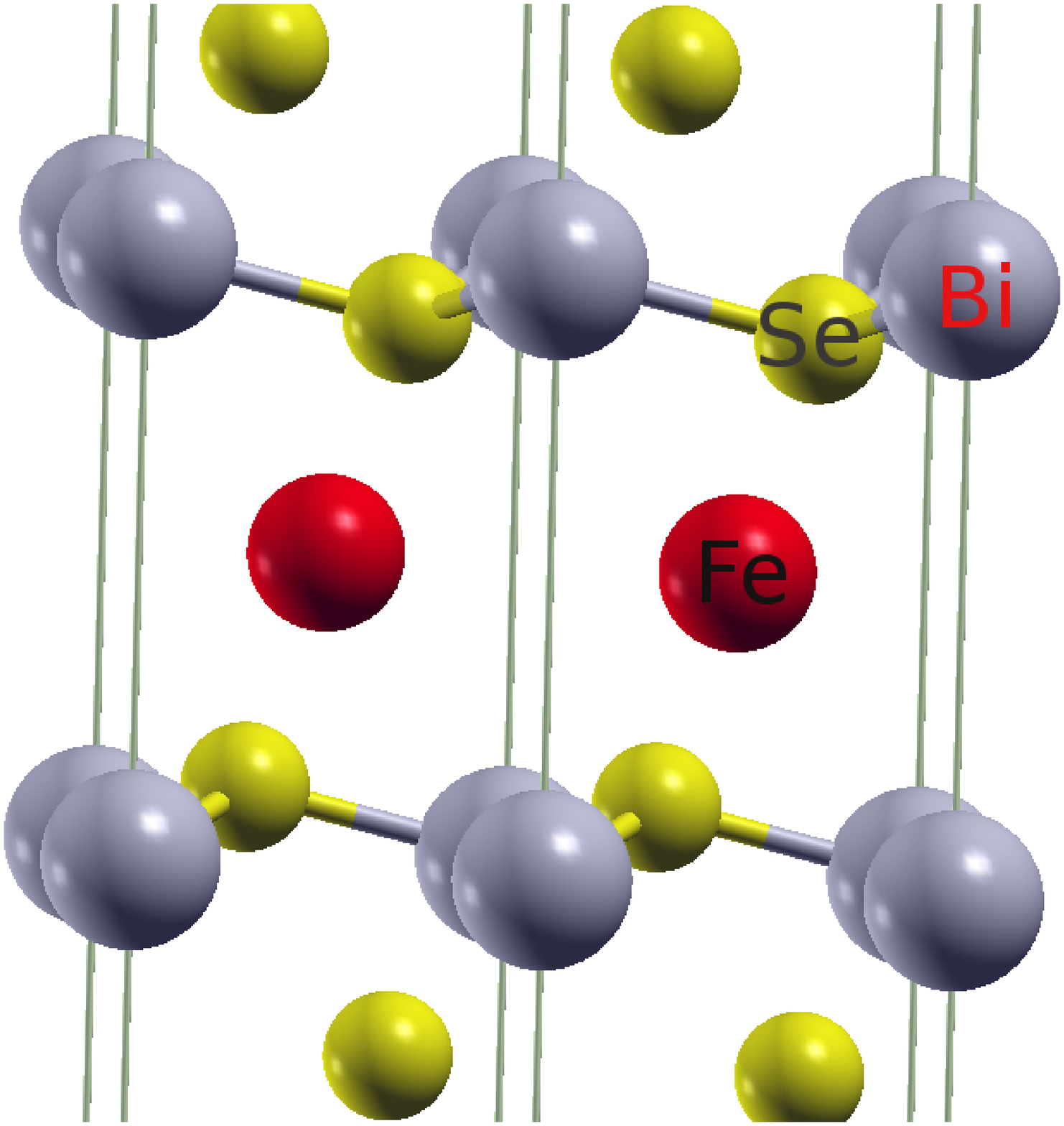}
(b)
\includegraphics[angle=0,width=0.45\columnwidth]{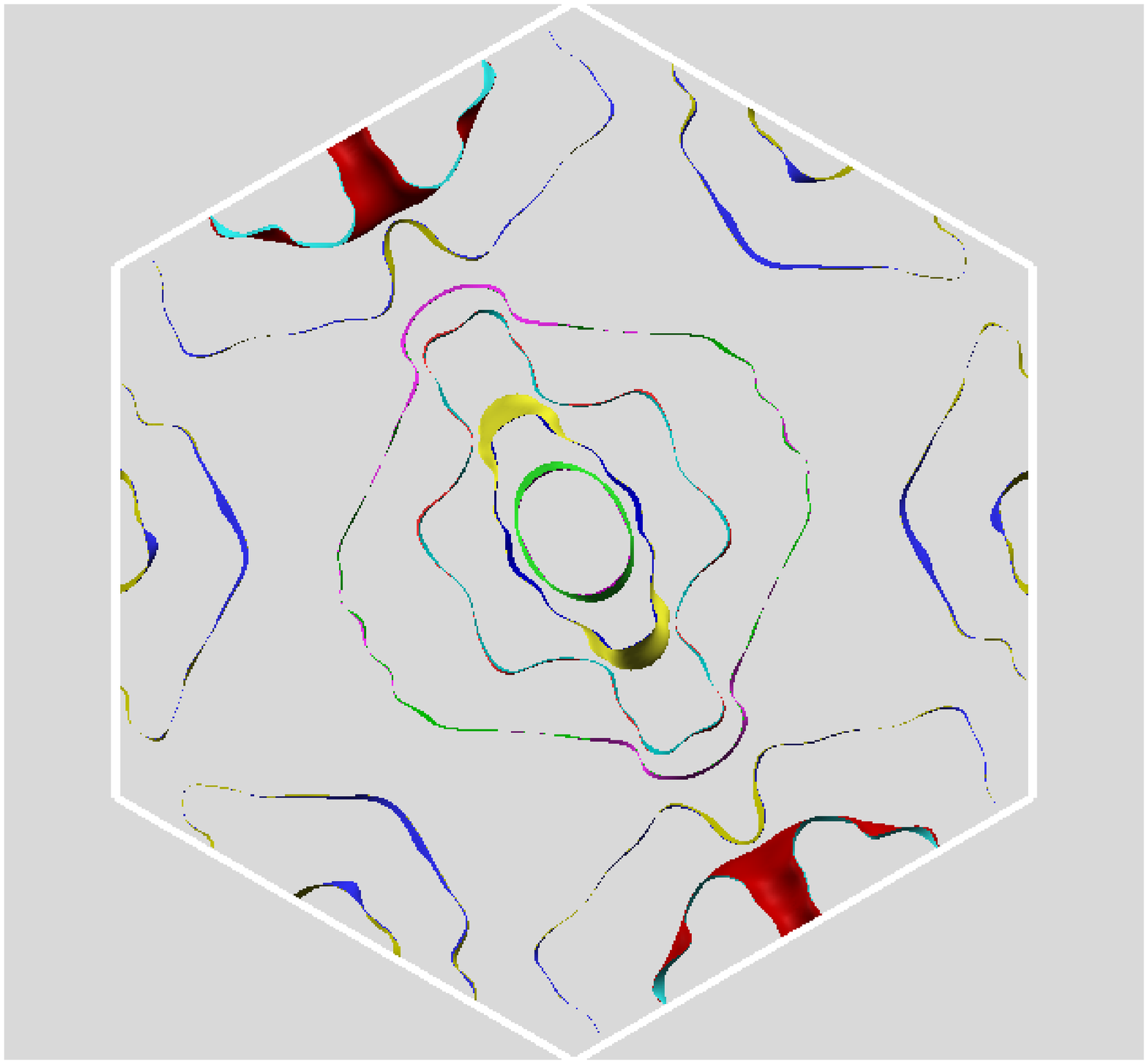}

(c)
\includegraphics[angle=0,width=0.45\columnwidth]{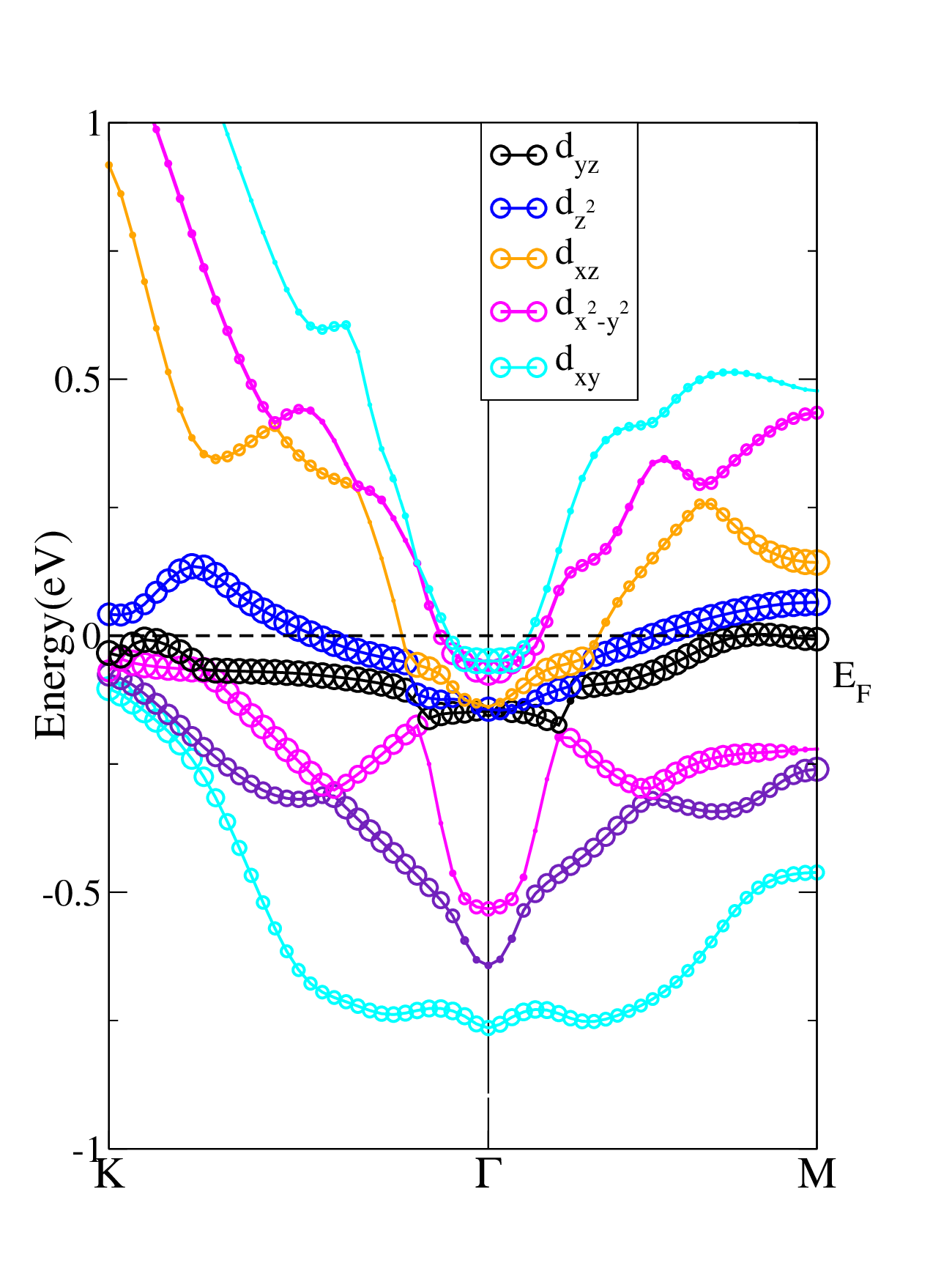}
\caption{(Color Online) (a) The crystal structure is shown where the atoms are labelled, 
(b) Fermi surface and (c) band structure of Fe intercalated Bi$_2$Se$_3$. The Fermi surface reveals 
impurity $d$ bands crossing the Fermi level. The high intensity region is for 
electrons, while the low intensity stands for holes.
Intercalation results in disappearance of the band inversion at 
the $\Gamma$ point and results in increase in the number of conduction 
electrons.} 
\label{fig1}
\end{figure}

\section{DFT+DMFT}
In this paper we use first-principles spin-polarized density functional 
theory (DFT) calculations combined with embedded dynamical mean field theory 
to demonstrate the intercalation induced changes in the electronic properties 
of Bi$_2$Se$_3$ using Fe as intercalant. First principles structure 
calculations were performed using WIEN2k full-potential linearized augmented 
plane wave (FP-LAPW) ab initio package\cite{pblaha} within the DFT \cite{dft} 
formalism to get the band structure and the density of states (DOS). A generalized gradient 
approximation Perdew-Burke-Ernzerhof (GGA-PBE) exchange correlation potential is
used here. The relativistic effects and the spin-orbit coupling  have been
included in the DFT calculation. In fact, for realization of the Dirac cone at the 
$\Gamma$ point in the parent Bi$_2$Se$_3$ crucially depends on the spin-orbit coupling.
The muffin-tin radii, R$_{MT}$ are chosen as 2.4 a.u. 
for Bi, 2.1 a.u. for Se and 2.3 a.u. for Fe atoms. The parameter, R$k_{max}$ (R$k_{max}$
stands for the product of the smallest atomic sphere radius R$_{MT}$ times
the largest $k$-vector $k_{max}$)
is chosen to be 7.0 and 1000 $k$-points with a 10 $\times$ 10 $\times$ 10 $k$-mesh is 
employed here for structural optimization. A $2\times2$ supercell is used for 
the intercalation of Fe. Intercalation is produced via introduction of the atoms
 into the interlayer space. The length of Fe-Fe bond in a layer is 0.45 nm 
while the length of Fe-Se bond along $c$ direction becomes 0.285 nm after 
intercalation. 
Further, for the total energy calculations, 
all atoms are relaxed until the maximum force is smaller than 0.01 $eV$/$\AA$.
Finally the self consistent field (scf) calculations are performed till an energy accuracy of  
0.0001 $eV$ is reached. The lattice constants obtained from our calculation agrees nicely 
with the previous experiment\cite{koski}.  
Then the band structure and the atom-resolved density of states are calculated from the
converged scf calculations (see Fig.1 and Fig.2). 

The spectral function shows partially occupied Se-4$p$, Bi-6$p$ and Fe-3$d$ 
bands near the Fermi level for the Fe-intercalated sample. The energy bandgap 
in parent Bi$_2$Se$_3$ 
is about 0.3 $eV$ (in good accord with the reported experiments and theory) closes with Fe-intercalation 
and impurity bands dominate near the Fermi level (see Fig.1c and Fig.2).
For performing correlated electronic structure and spectra calculations,
a fully charge-self-consistent dynamical mean field theory (DMFT) is employed via the  
EDMFTF package\cite{haule}, which implements a combined DFT and DMFT derived 
from the stationary Luttinger-Ward functional. Here, the exact double-counting 
of DFT and DMFT and Coulomb interaction are well treated and the Green's 
function is determined self-consistently. In strongly correlated systems, DFT+DMFT has been successful 
in detailing a lot of important results\cite{kim,eug,at1,su,kotliarrmp,supc}. In the DMFT section, the  
non-interacting Hamiltonian is added with the Coulomb interaction term, 
H$_{int}$\cite{aginter}, to incorporate the effects of correlated Fe-3$d$ orbitals and also a self energy functional,
$\Sigma_{dc}$ to take care of the double counting. 
The total Hamiltonian, except the $\Sigma_{dc}$ term is expressed as, 
\begin{equation}
H=\sum_{k,a,\sigma}\epsilon_{k,a}c^{\dagger}_{k,a,\sigma}c_{k,a,\sigma}+U\sum_{i,a}n_{ia\uparrow}n_{ia\downarrow} + $$
$$U'\sum_{i,a,b,\sigma,\sigma'}n_{ia\sigma}n_{ib\sigma'}-J_H\sum_{i,a,b}S_{ia}.S_{ib}
\end{equation}
where $\epsilon_{k,a}$ is the band dispersion which includes the
effects of SOC, $\sigma$ stands for up and down spins and 
$U$ and $U'$ are the intra- and inter-orbital Coulomb interaction terms between electrons with 
opposite spins in the same orbital and between electrons 
with parallel spins in different orbitals respectively, and $J_H$ is
the Hund’s coupling. The inter-orbital term ($U'$) is reduced by
the ferromagnetic coupling due to Hund’s first rule that favors the alignment of spins. 
We also have a relationship between different energy scales, namely, $U$, U$'$ and $J_{H}$ given by,
$U'=U-2J_{H}$. In our work we have considered $J_H$=1.25 eV (reasonable for Fe-3$d$ bands) and varied 
$U$ over a realistic range. 

Finally the total Hamiltonian is solved 
using the DMFT method. The correlated five Fe-3$d$ orbitals are treated dynamically
within the DMFT based on orbital projection-embedding scheme accomplished via the
EDMFT package, while the Bi and Se-$p$ orbitals are treated at the DFT level. The 
impurity solver used in the DMFT code is the continuous time 
quantum Monte Carlo(CT-QMC) in the hybridization expansion 
method \cite{haule1}. The parameters, namely, the Coulomb interaction $U$, Hund's coupling,
($J_H$) and the inverse temperature, $\beta$ ($ = 1/k_{B}T$) are varied within an experimentally realizable range to get 
$T$ and $U$ dependence of the intercalated system. The DFT+DMFT calculations are
converged upto presision of 0.0001 with respect to the charge density, the 
chemical potential and the self energy with considering step over E as $10^{-6}$. 
Finally the maximum entropy method\cite{jarrell} is used for the 
analytical continuation of the self-energy from the imaginary axis to real 
frequencies with an auxiliary Green's function. Then from the real frequency 
Green's function, the momentum-resolved spectral functions and the density of states are
obtained. To check the stability and accuracy of the result we have used Pade 
approximation in addition to the maximum entropy method used in the EDMFTF package.      

\begin{figure}
\includegraphics[angle=270,width=0.8\columnwidth]{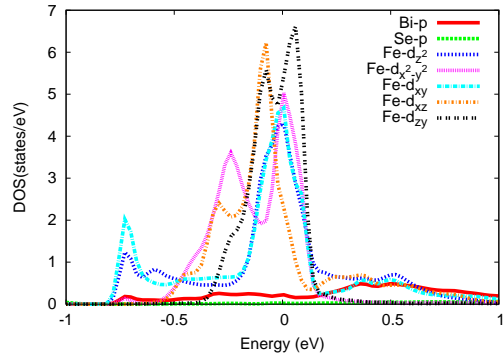}
\caption{(Color Online) Spectral function of Fe intercalated Bi$_2$Se$_3$. For the effective model incorporating conduction electrons within 
DMFT calculations, we use the Bi-p, Se-p and Fe-3d bands crossing 
Fermi Energy ($E_F$=0), while Bi-p and Se-p bands have very less spectral weights
at the Fermi energy, the Fe-$d$ bands dominate the low energy physics.}
\label{fig2}
\end{figure}
\begin{figure*}
(a)
\includegraphics[angle=270,width=0.9\columnwidth]{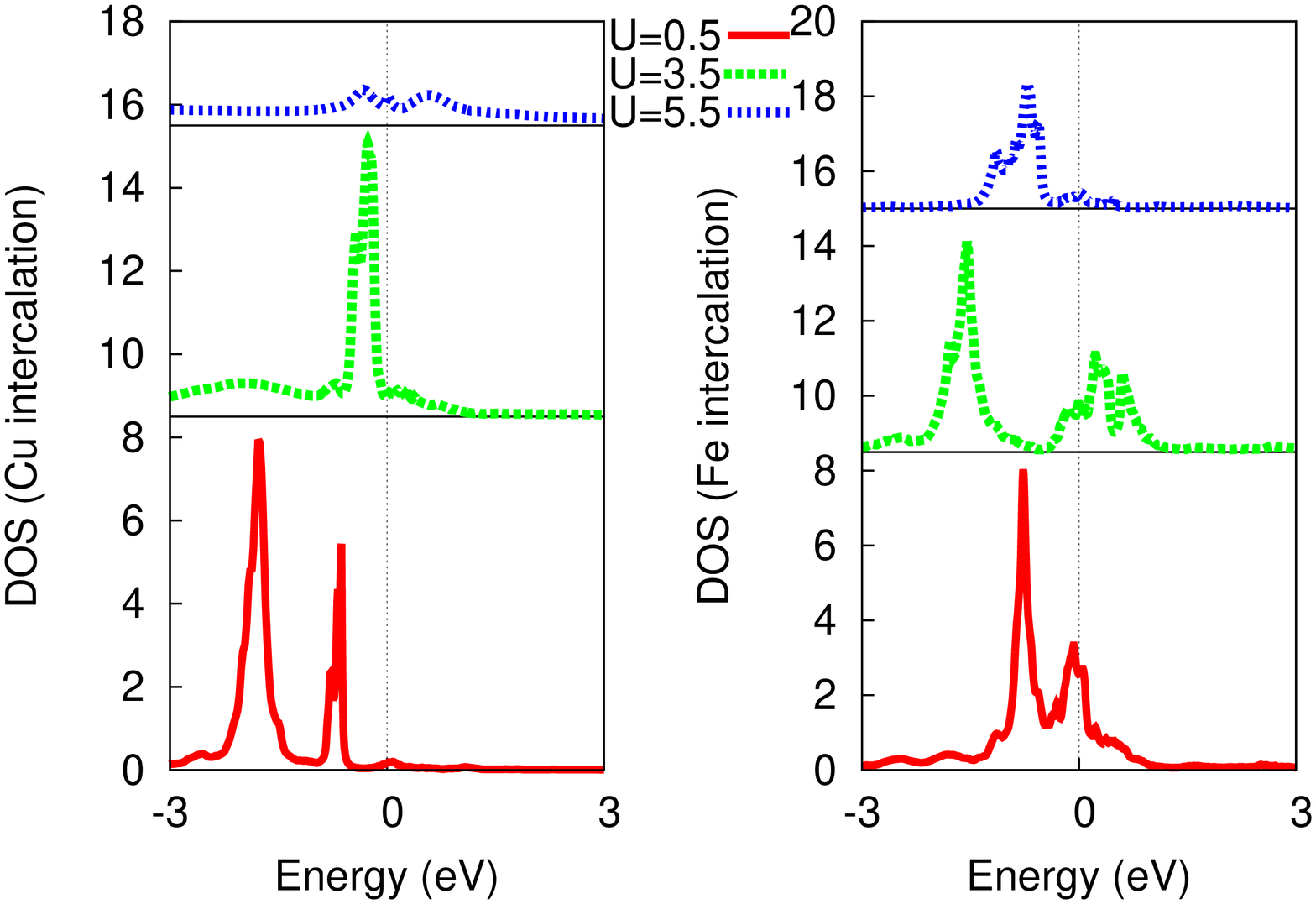}
(b)
\includegraphics[angle=270,width=0.9\columnwidth]{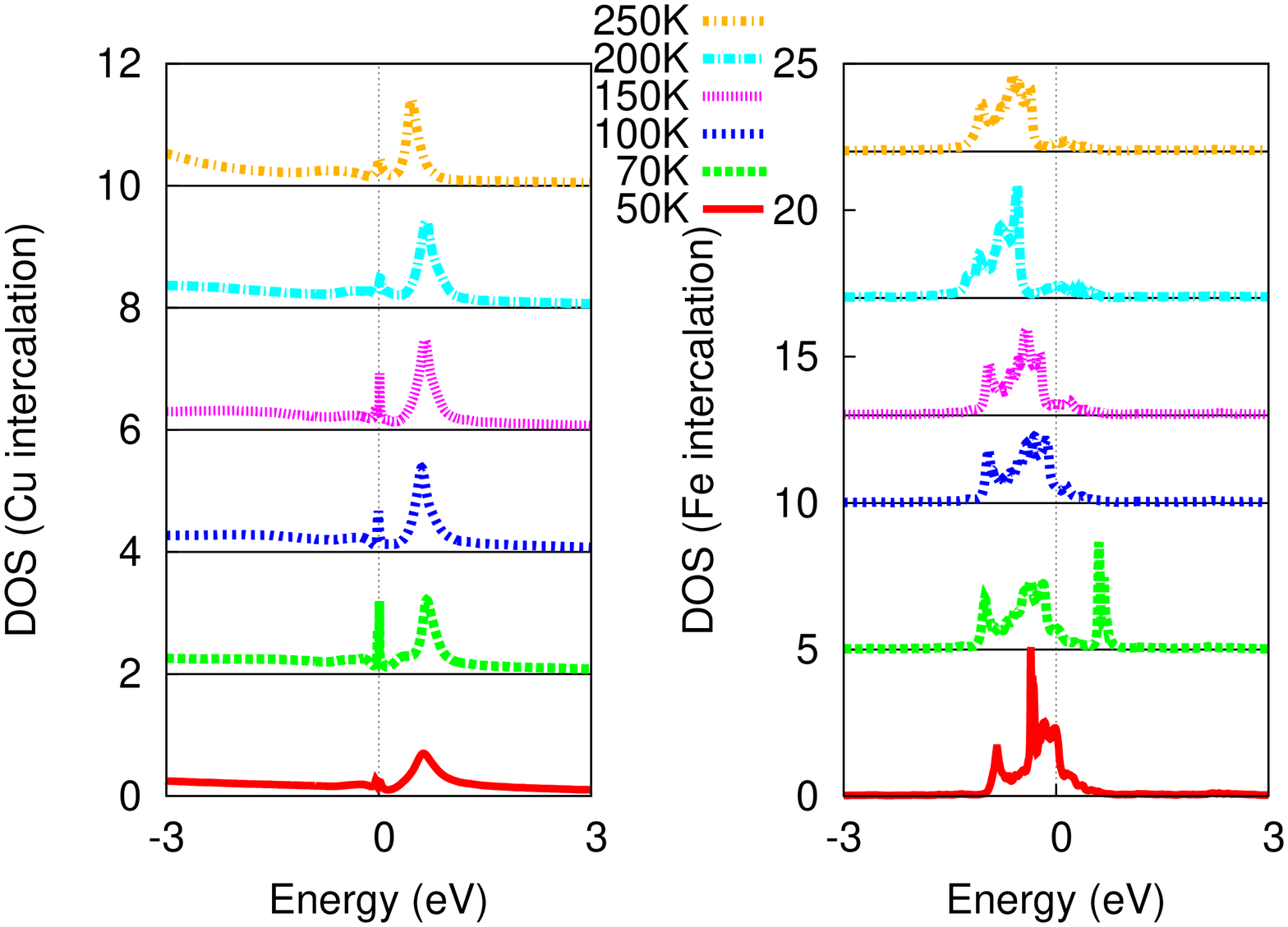}
\caption{(Color Online) Total density of states 
from DFT+DMFT for Cu and Fe intercalated Bi$_2$Se$_3$ at
(a) various values of $U$ and (b) at different temperatures, $T$. DFT+DMFT
density of states for Cu intercalation is provided to compare with the earlier 
results\cite{sasaki}. Spectral density at $E_F$ remains finite throughout the temperature range
for Cu intercalation, whereas for the Fe-intercalated compound, it decreases with 
temperature and the system behaves as correlated bad metal. Density of states are shifted 
along the $y$-axis to reflect changes induced by Coulomb interaction and temperature respectively.}
\label{fig3}
\end{figure*}

\section{Results and Discussion}

We have considered Fe intercalation in between two quintuple layers of Bi$_2$Se$_3$, a
situation that is energetically more favorable. In each of the
quintuple layer, the hexagonal atomic planes are arranged following the 
sequence of Se1-Bi-Se2-Bi-Se1 along the $z$-direction with covalent 
bonding between the atoms, and the Se1 and Se2 atoms stand for two inequivalent 
Selenium atoms. So the intercalated Fe atom will be in the environment of a Se1 
atom. First we study the change in the band structure of Bi$_2$Se$_3$
due to the intercalation by Fe (see Fig.1c).

The validity of the EDMFT package is well established \cite{eug,kim}. 
Moreover our calculated results, such as the momentum resolved spectral function 
and the magnetic order are consistent with the earlier iron doped Bi$_2$Se$_3$ results\cite{jung}. We also present DMFT 
one particle spectrum for Cu intercalated Bi$_2$Se$_3$ which shows good 
agreement with the earlier experiments\cite{sasaki} (zero energy peak in the density of states). 
Since a Fe-based system consists of unfilled 3$d$-orbitals and possesses 
different values of the Coulomb interaction, we varied it within a reasonable range,  
and determined the final values of $U$ and $J_H$ for the system (considering earlier 
experimental results), which are found as 5.5 eV and 1.25 eV respectively. 
Large values of $U$, that is, strong electronic correlations renormalize the 
spectral function considerably. The total density of states of an intercalated 
Bi$_2$Se$_3$ is presented in Fig.3 for different values of $U$. 
In comparison with the DFT DOS, the correlated electronic spectra from the DMFT are
very distinct, indicating that this system is a correlated bad metal. It is 
observed that, the spectral weight at the Fermi level, $E_F$ reduces with increasing $U$. 
The Fe intercalation leads to an abrupt change of the electronic structure 
and the 3$d$ orbitals of the Fe atom dominate the low-energy properties of the compound (Fig.2). 
This is in contrast with Fe-doped Bi$_2$Se$_3$ which shows an insulating behaviour,
and instead here we get bad metal. Further, we obtain the temperature dependence of the DOS,
by employing a different value of $\beta$ in the EDMFTF. This yields that with decreasing temperature,
the total spectral weight of the $d$-orbital DOS increases at the Fermi level, while at around 50 K,
the $d_{z^2}$ and $d_{x^2-y^2}$ orbitals undergo an orbital selective metal-insulator 
transition as shown in Fig.4a. The orbital selectivity is confirmed from the
divergence of Im$\Sigma(\omega)$ at the Fermi level for the $d_{z^2}$ 
and $d_{x^2-y^2}$ orbitals, while the other orbitals provide support for 
metallic features at low energies. This orbital selectivity  at 50K can be 
related to the onset of a possible magnetic order, as also corroborated by the 
magnetic susceptibility plot (Fig.4b). The right inset of Fig.4a shows Im$\Sigma(\omega)$ 
at 100 K which reveals non-Fermi liquid features due to high $T$, while the left 
inset of Fig.4a depicts the resitivity data at low temperatures, which shows proportionality to $T$
with a change of slope around 50 K. Change of slope around 50 K in resistivity 
further confirms an ordering transition to occur. This ordering transition 
is also coherence restoring transition because resistivity decreases at lower 
$T$. The transport observation also goes hand in hand with our 
prediction of a paramagnetic to a ferromagnetic transition, while the paramagnetic 
order can be considered as fluctuation of magnetic moments at high $T$.
The observation of orbital selectivity and related magnetic order in 
Fe-intercalated systems
is not surprising and can be argued to have originated from moderate $U'$ ($U'$ being
the inter-orbital Coulomb repulsion) which leads to a selective Mottness in the multi-band
situation. Interestingly, stabilization of the magnetic order enhances selective
Mott features at low temperatures as a zero energy pole emerges in the 
Im$\Sigma(\omega)$ ($d_z^2$ and $d_{x^2-y^2}$, along with small mass 
enhancement in Re$\Sigma(\omega)$ of three other d-orbitals.

\begin{figure}
(a)
\includegraphics[angle=270,width=0.7\columnwidth]{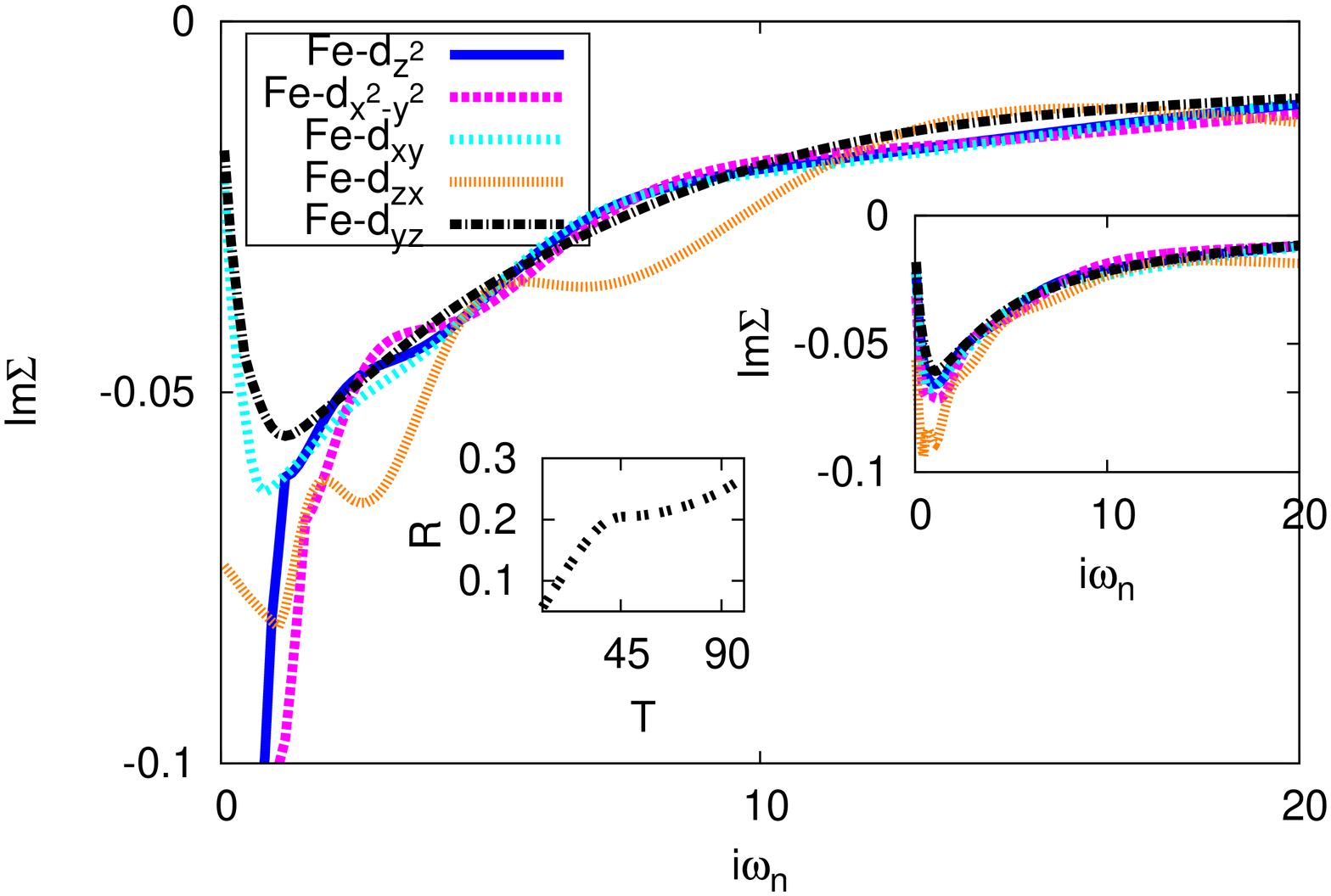}

(b)
\includegraphics[angle=270,width=0.7\columnwidth]{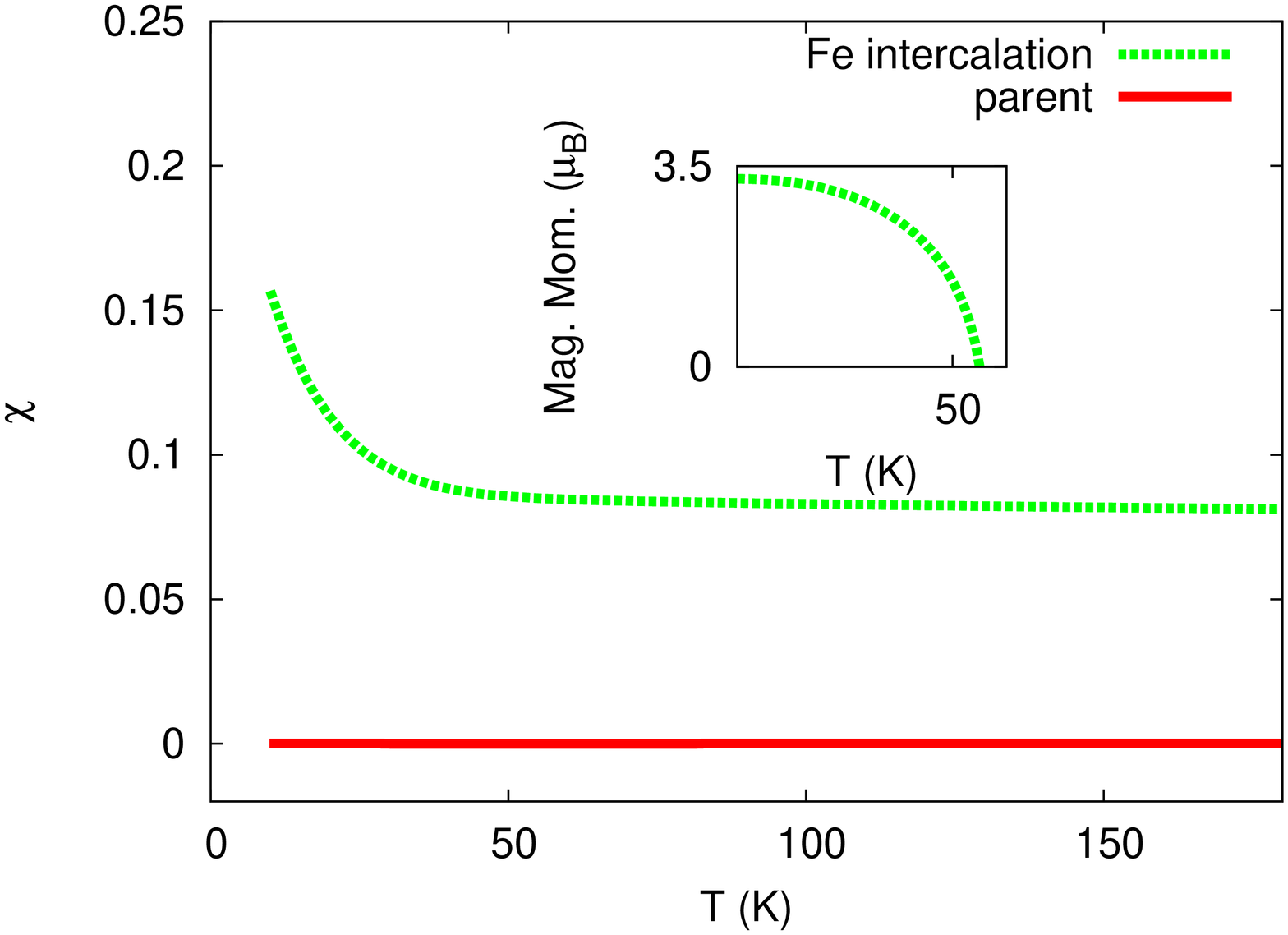}
\caption{(Color Online) (a) Orbital dependent self energies from DFT+DMFT for 
Fe intercalated Bi$_2$Se$_3$ at combination of $U$, $U'$ (5.5 eV and 3.0 eV) and 
$T$= 50 K and $T$= 100 K (right inset Fig.4a). Left inset shows normalized 
resistivity with temperature.
(b) $T$ dependent DMFT susceptibility, $\chi$ for the parent and the intercalated Bi$_2$Se$_3$. 
$\chi$ for the intercalated compound shows steep increase at around 50 K, while as shown
in the inset, the magnetic moment, $m$ vanishes around the same temperature.}
\label{fig4}
\end{figure}
\begin{figure}
(a)
\includegraphics[angle=0,width=0.7\columnwidth]{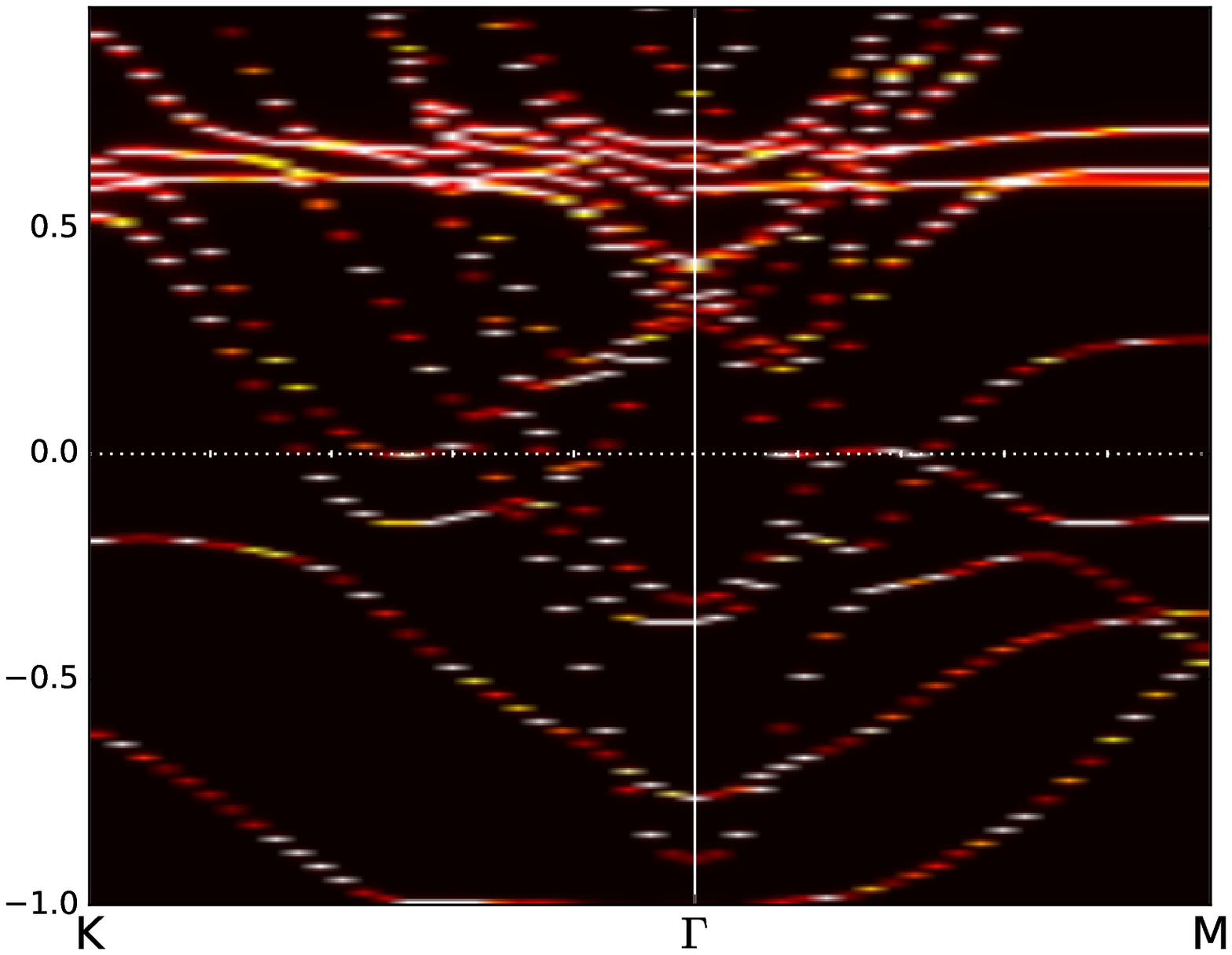}

(b)
\includegraphics[angle=0,width=0.7\columnwidth]{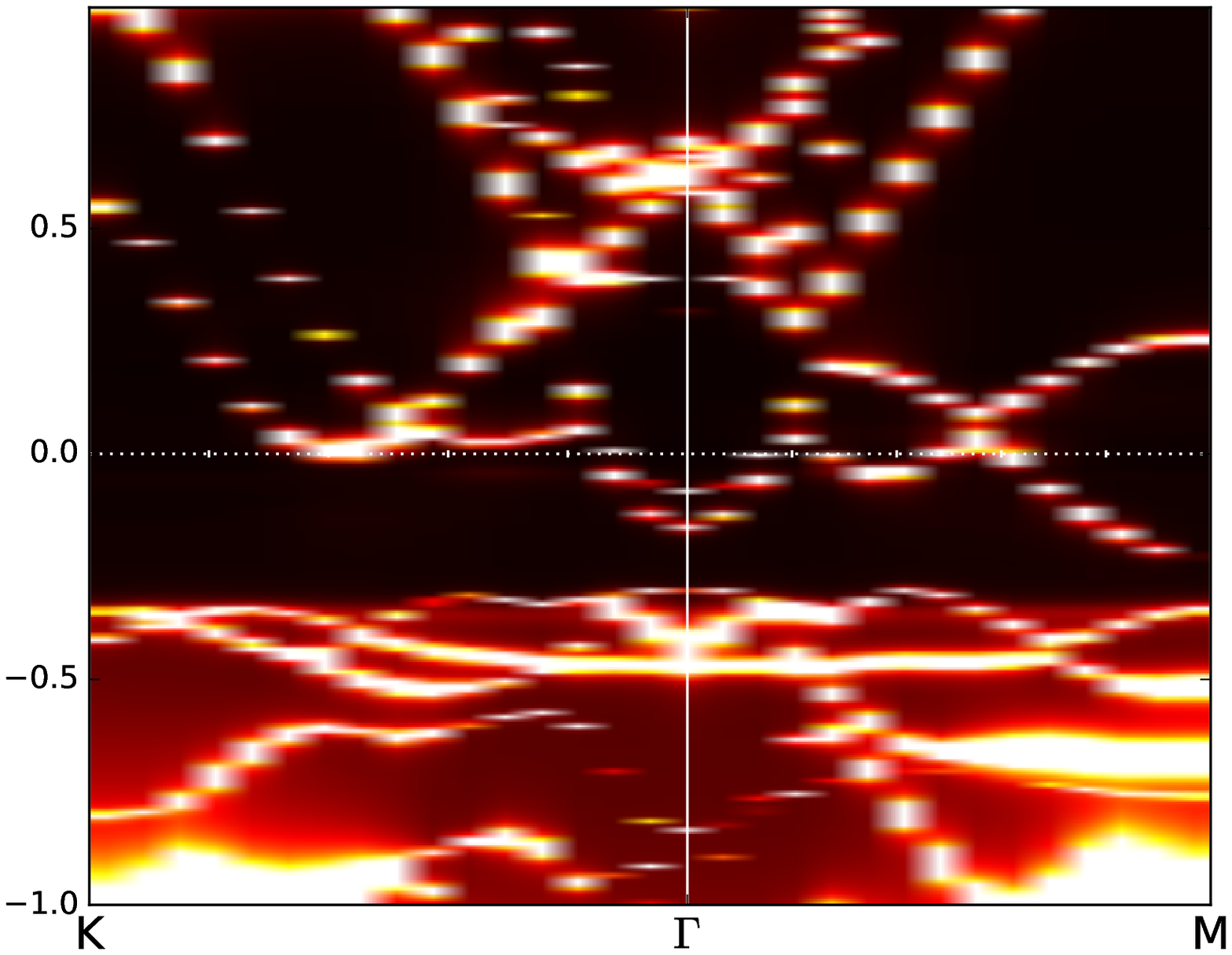}
\caption{(Color Online) Momentum resolved spectral function map for Fe 
intercalated Bi$_2$Se$_3$ at (i) low (50K) and (ii) high (200 K) temperatures.}
\label{fig5}
\end{figure}

\begin{figure}
(a)
\includegraphics[angle=0,width=0.4\columnwidth]{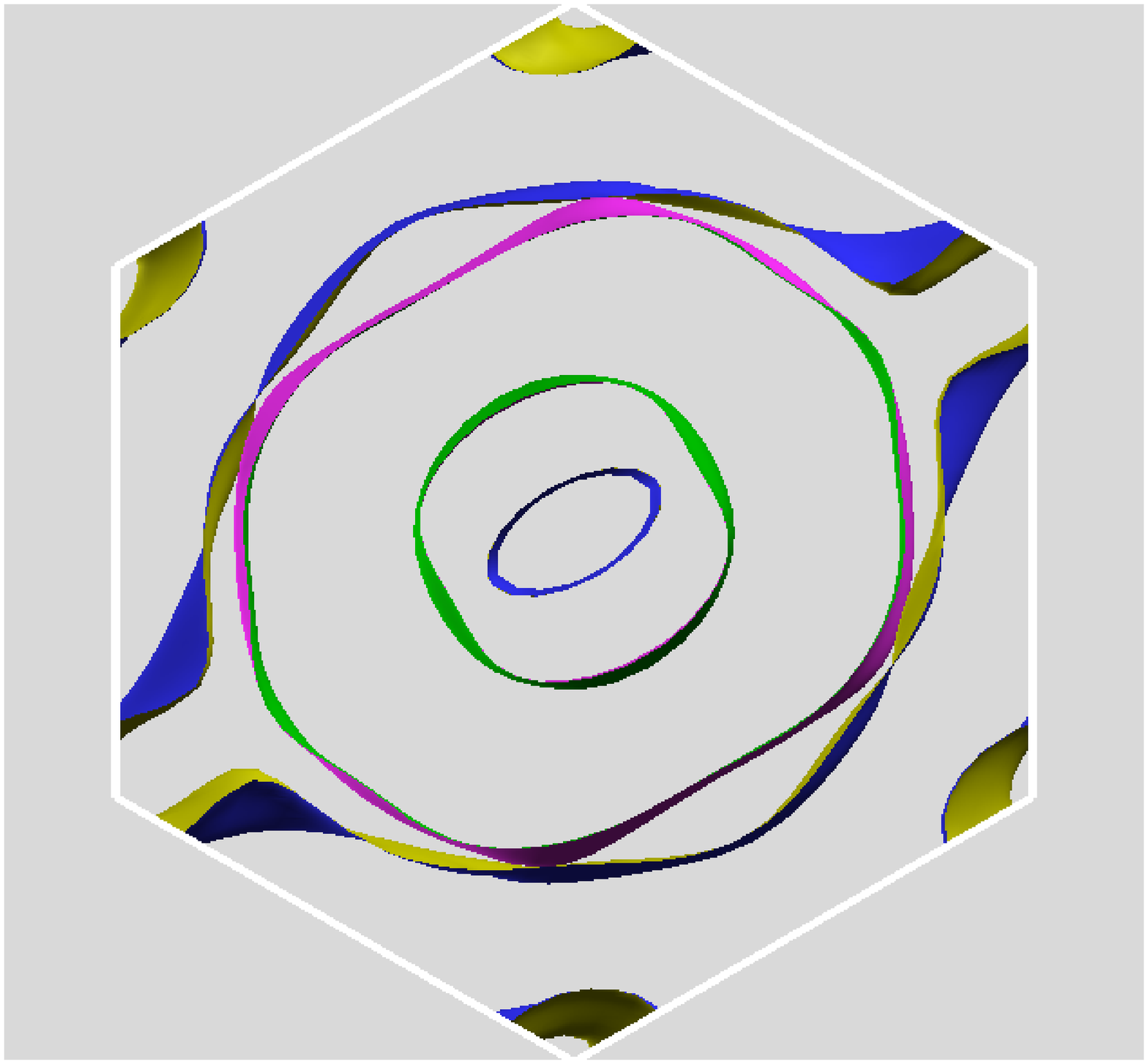}
(b)
\includegraphics[angle=0,width=0.4\columnwidth]{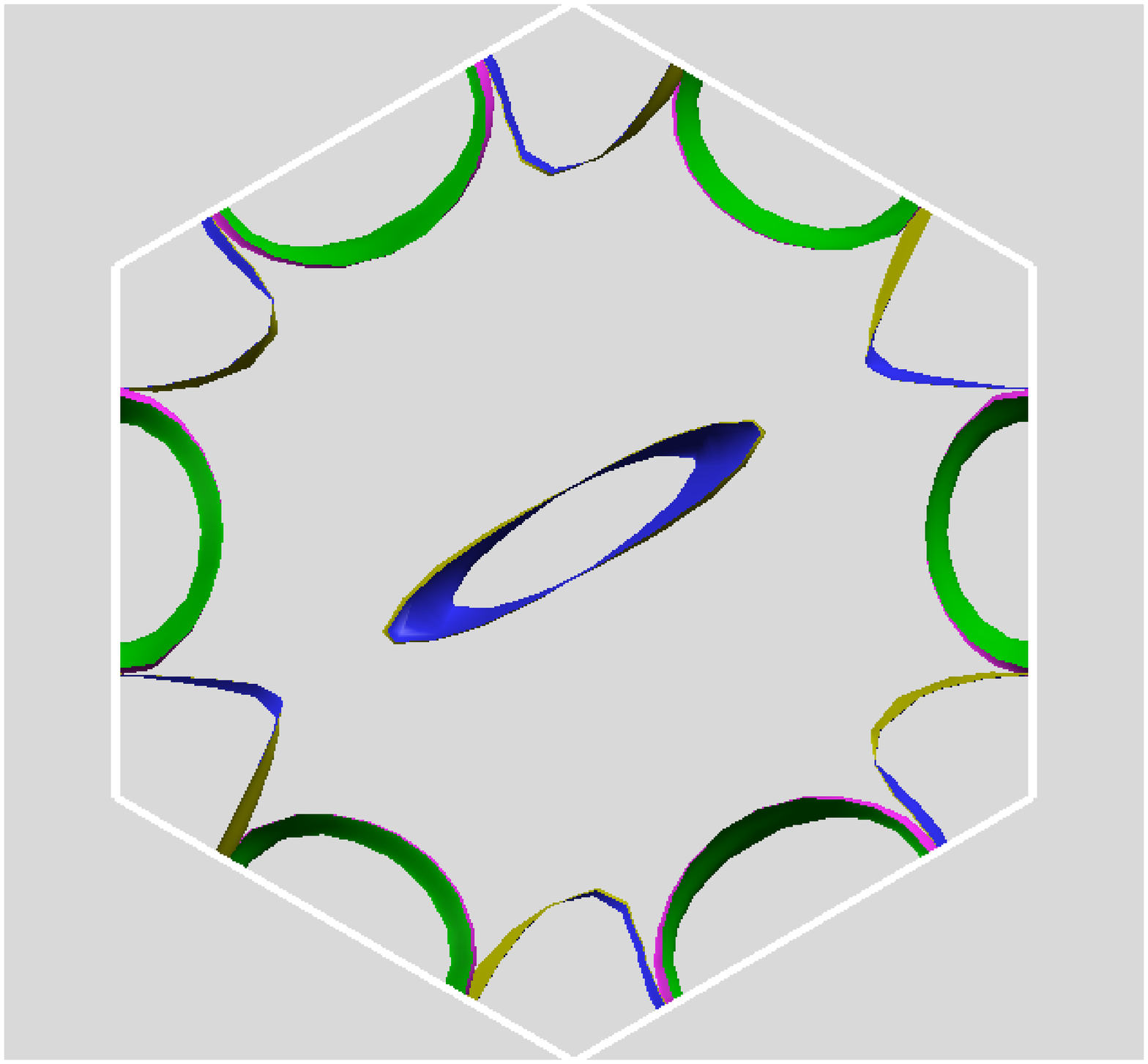}
\caption{(Color Online) Fermi surface map for Fe 
intercalated Bi$_2$Se$_3$ at (i) low (50K) and (ii) high (200 K) temperatures. 
Emergence of new pockets in the Brillouin zone edge is evident, whereas none 
of the electron pockets around the $\Gamma$ point disappear.}
\label{fig6}
\end{figure}

Next we present angle resolved photoemission spectra of Fe intercalated 
Bi$_2$Se$_3$ in Fig.5a and Fig.5b. In contrast to the conventional DFT 
band structure results, it shows pronounced renormalized effects in the 
band structures owing to strong electronic correlations. The characteristic 
Dirac-like dispersion manifests above the Fermi level at the $\Gamma$ point for both 
low (50 K) and high (200 K) temperatures. Near the Fermi level (FL), two bands cross the
FL along the $K-\Gamma-M$ direction. Further at the lower temperature (that is, 50 K), the lower band sinks below 
the FL signaling a transition from a hole pocket to an electron pocket in the 
$\Gamma-K$ and $\Gamma-M$ directions. In comparison with the DFT band structure 
results, the DMFT momentum resolved spectra indicate a correlation driven 
Lifshitz transition. It may also be noted that in this study, the intercalation density is $10\%$ 
(we have examined other (lower) densities, although not presented here), as 
this scenario seems to be ideal for observing a magnetic phase transition. 
Our results also indicate emergence of small electron pockets around these 
two directions (namely, $\Gamma-K$ and $\Gamma-M$), which is further confirmed by the DMFT Fermi
surface (FS) results (see Fig.6) discussed below.

Further to address intercalation dependent changes we calculate the DMFT FS
at two different temperatures (namely, 50 K and 200 K). Combined with intercalation
dependent evolution of the electronic structure, the Fermi surface map reveals 
connection between ordering and electronic correlations. Since the effect of 
electronic correlations is already described above, the DMFT FS deserves 
to be investigated. Both the DFT FS (Fig.1b) and the DMFT FS (Fig.6) are shown here for 
comparison. Both the results reveal presence of electron pockets around the
$\Gamma$ point and hole pockets around the $K$ and the $M$ points. In the DMFT results, with
strong correlation being considered, the FS topology shows a sizeable change, that is, 
from five electron pockets, it becomes three electron pockets at 50 K and finally a one 
electron pocket at 200 K around the $\Gamma$ point. On the other hand, the hole pockets 
around the $K$ and $M$ points also undergo significant modification. We strongly believe that 
these modifications arise due to orbital dependent electronic structure 
reconstruction and essentially driving the system into a magnetically ordered phase.

To capture the nature of magnetic ordering by electronic correlations, we have
calculated the temperature variation of the magnetic susceptibility and the moment from the EDMFT package.  
In Fig.4b we show the temperature dependence of the magnetic susceptibility, $\chi$ and the inset contains
the variation of the magnetic moment, $m$ with temperature. 
The parent compound Bi$_2$Se$_3$ possesses a weak diamagnetic character with almost
no temperature dependence, while Fe intercalation makes the compound magnetic.
The magnetic nature is confirmed to be ferromagnetic as $\chi$ shows a Curie behaviour with
$\chi \sim (T-T_c)^{-1}$, with $T_c$ as the Curie temperature. We have observed $T_{c} \sim$ 30K.
Moreover we have calculated the spin gap energy (denoted by the 
energy difference between the ferromagnetic and the weakly diamagnetic state), 
which reveals that the ferromagnetic state is more energetically stable. In 
Fe-Bi$_2$Se$_3$ system,
the magnetic moments of Fe is about 3.5$\mu_B$ and the other atoms have smaller magnetic moments (of about 
0.001-0.01$\mu_B$) which align antiparallely with the moments of the Fe atom. 
The system turns out to be magnetic with a value for the magnetic 
moment as 3.4$\mu_B$. The spin up and spin down states are partially filled 
with a occupation difference between them resulting in a magnetic moment 
which is almost totally
contributed by the Fe-3$d$ orbitals and the relevant electronic correlations.

In conclusion, the electronic correlations along with a reasonable spin-orbit 
coupling in Fe intercalated Bi$_2$Se$_3$ have been examined employing the 
DFT+DMFT method executed with the CT-QMC impurity solver. 
A strong correlation-driven electronic structure modification accompanied by a Lifshitz transition 
is found in the intercalated compound, indicating an ordered phase to set in at lower
temperatures, which is significantly different than the parent compound. Both the electronic correlations
and the spin-orbit coupling play major roles in the electronic structure and 
transport properties of intercalated compound. These generate tiny electron 
pockets around the $\Gamma -K$ and the $\Gamma -M$ directions. Our results are somewhat similar
to the Fe-doped (not intercalated) Bi$_2$Se$_3$ compound, in the sense that it also shows different type of 
magnetic ordering with doping concentration and temperature\cite{salman}. 
All these results are new in literature and subject to further experimental 
study. The results mainly demonstrate many-body characteristics, such as 
many-body self-energy, spectral weight, presence of quasiparticles 
 near the Fermi level, and a strong renormalization of the effective masses. 
Further explorations of the low temperature properties and other intercalation 
densities need to be undertaken in near future. Moreover increased $T_c$ due to intercalation 
can lead to potential applications in the context of spintronic devices and energy 
storage materials\cite{ref7-10koski,11,12,13}.  Our studies suggest that the 
Fe-intercalated Bi$_2$Se$_3$ could be an ideal system to identify the role of 
electronic correlations and spin-orbit coupling in magnetism of iron-based 
materials. 
\acknowledgements
This work is financially supported by DST women scientist grant 
SR/WOS-A/PM-80/2016(G). S. K. gratefully acknowledge Prof. M C Mahato for mentoring and useful conversations.
SB acknowledges support from the SERB grant EMR/2015/001039.

\end{document}